\documentclass[a4paper,twocolumn,11pt,accepted=2026-08-03]{quantumarticle}
\pdfoutput=1
\usepackage{graphicx} 
\usepackage{dcolumn} 
\usepackage{subcaption}
\usepackage{amssymb}
\usepackage{amsmath}
\usepackage{amsfonts}
\usepackage{mathbbol}
\usepackage{latexsym}
\usepackage{theorem}
\usepackage{hyperref}
\usepackage{float}
\usepackage{xcolor}
\usepackage{lipsum}
\usepackage{microtype}
\usepackage[sort&compress,numbers]{natbib}
\usepackage[utf8]{inputenc}
\usepackage{bm}
\usepackage{verbatim}
\usepackage{mathrsfs}
\usepackage{color}
\usepackage{dsfont}
\usepackage{booktabs}

\newcommand{\ket}[1]{|#1\rangle}
\newcommand{\bra}[1]{\langle #1 |}

\begin{document}

\title{Disorder-Induced Entanglement Phase Transitions in Non-Hermitian Systems with Skin Effects}

\author{Kai Li}
\affiliation{Center for Quantum Information, IIIS, Tsinghua University, Beijing 100084, People's Republic of China}
\thanks{These authors contribute equally to this work.}

\author{Ze-Chuan Liu}
\affiliation{Center for Quantum Information, IIIS, Tsinghua University, Beijing 100084, People's Republic of China}
\thanks{These authors contribute equally to this work.}

\author{Yong Xu}
\email{yongxuphy@tsinghua.edu.cn}
\affiliation{Center for Quantum Information, IIIS, Tsinghua University, Beijing 100084, People's Republic of China}
\affiliation{Hefei National Laboratory, Hefei 230088, PR China}

\begin{abstract}
Non-Hermitian dynamics is ubiquitous in various physical systems.
While recent study shows that such a dynamics leads to an area-law scaling of the entanglement entropy due to the non-Hermitian skin effects,
it remains unclear how disorder changes the behavior of the entanglement entropy in a non-Hermitian system with skin effects.
Here we study the dynamics of a many-body state of free fermions in the paradigmatic Hatano-Nelson model with open boundaries,
and find that the area-law behavior of the entanglement entropy in the pristine Hatano-Nelson model
develops into a logarithmic scaling for small disorder strength. As we further increase the disorder strength, the system reenters an area-law regime through an entanglement phase transition.
At the critical point, the entanglement entropy exhibits a universal algebraic scaling.
We further demonstrate the absence of a conformal invariance in the log-law regime by examining the subsystem entanglement entropy, the
connected correlation function and the mutual information. Finally, we show the existence of disorder induced entanglement phase transitions in the Hatano-Nelson model with periodic boundaries.
\end{abstract}
\maketitle

Non-Hermitian physics has received considerable attention in the past few years due to the discovery
of various peculiar phenomena~\cite{ChristodoulidesNPReview,XuReview,ZhuReview,UedaReview,BergholtzReview}, such as
non-Hermitian intrinsic topology~\cite{TonyLee,Xu2017PRL,Nori2017PRL,Fu2018PRL,Ueda2018PRX,Kawabata2019PRX, Hengyun2019PRB, Xu2022PRL, Xu2023PRL}
and non-Hermitian skin effects~\cite{Yao2018PRL1,Xiong2018JPC,Torres2018PRB,Kunst2018PRL,Okuma2020PRL,Slager2020PRL,ChenFang2020PRL,Qibo2020}.
Moreover, non-Hermiticity is prevalent in the dynamics of quantum systems~\cite{QMCbook}.
In fact, when continuous measurements and postselection are considered, the dynamics of a many-body system is governed by a non-Hermitian Hamiltonian~\cite{QMCbook}.
In this context, interesting phenomena have been found in non-Hermitian entanglement dynamics~\cite{
Ueda2018PRL, Ueda2019PRL, Lucas2020PRR, Balazs2021PRB, Gullans2021PRL, Schiro2021PRB, Schiro2021Quantum, Chen2021PRB, Imura2022PRB, Ryu2022arXiv, Schiro2023PRB, Schiro2023scipost, Pal2023arxiv, Hamazaki2023arxiv, Imura2023PRB},
including entanglement and purification transitions~\cite{Gullans2021PRL}, as well as skin effects-induced entanglement phase transition~\cite{Ryu2022arXiv}.

Disorder plays a crucial role in the behavior of physical systems.
It is widely known that disorder can induce Anderson localization~\cite{Anderson1958PR}, which modifies the transport properties~\cite{Ramakrishnan1985RMP, Mirlin2008RMP}
and also suppress the growth of entanglement~\cite{Huse2015}.
Within the scope of non-Hermitian systems, the interplay between non-Hermiticity and disorder can result in unique properties~\cite{UedaReview}.
For example, the transition between skin states and Anderson localized states induced
by disorder has been discovered~\cite{Chen2019PRB, Longhi2019PRL, Szameit2022Nature, Ryu2021PRL}.
Furthermore, disordered non-Hermitian systems can exhibit the generalized mobility edge, separating extended and localized states in the complex energy
plane~\cite{HN1996PRL, HN1998PRB, Qibo2020PRR, ChenPRB2020, Longhi2020PRB, LWLZC2021PRB}.
In the context of dynamics, the presence of skin effects can inhibit the entanglement growth,
causing an area-law entanglement for the steady-state in disorder-free systems~\cite{Ryu2022arXiv}.
One may expect the introduction of disorder in such a system could result in an entanglement phase transition corresponding to the change of localization properties.
However, it remains unclear whether this phase transition would exhibit similar properties to those observed in Hermitian systems, or if distinct critical behaviors would emerge.

\begin{figure}[t]
\centering
\includegraphics[width=8.3cm]{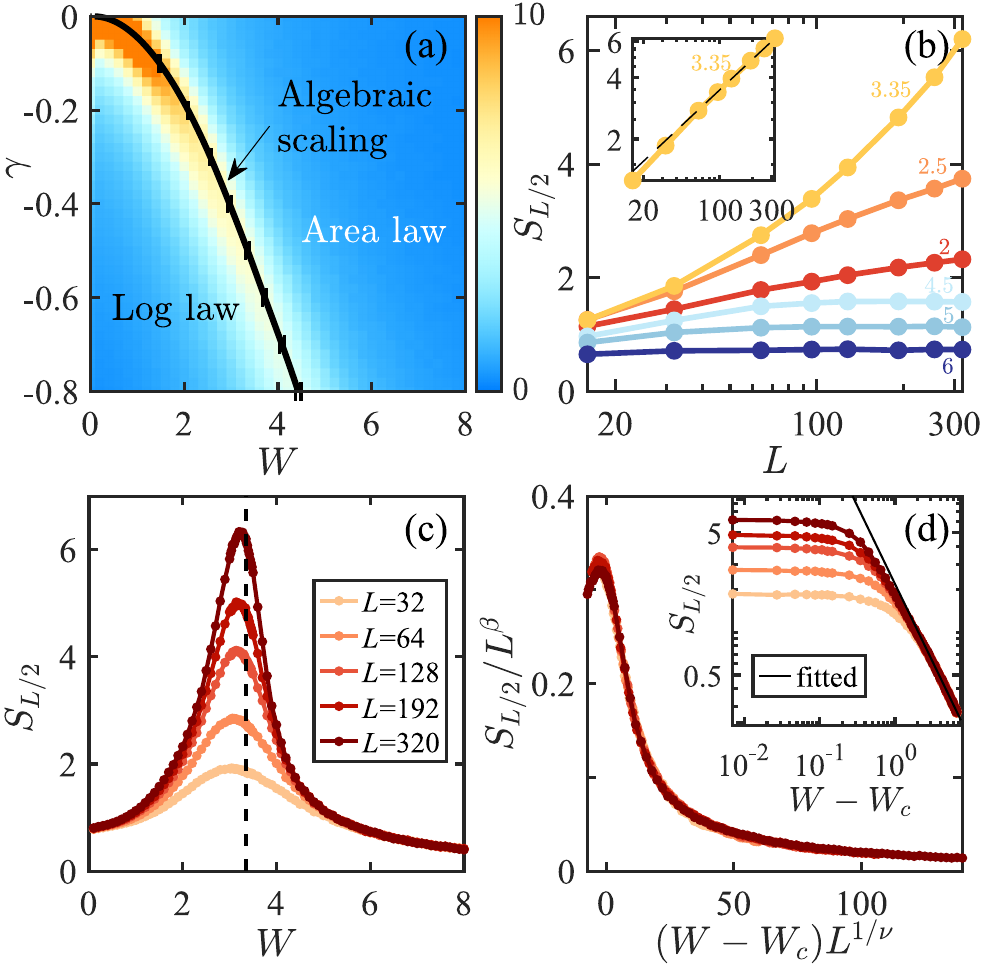}
\caption{(a) The phase diagram of the HN model in Eq.~(\ref{Disordered_HN_Model}) with respect to the disorder strength $W$ and the asymmetric hopping strength $\gamma$.
Color denotes the entanglement entropy $S_{L/2}$ of a half system for $L=256$.
The black line describes the phase boundary obtained by finite-size scaling of the entanglement entropy.
(b) The linear-log plot of $S_{L/2}$ versus the system size $L$ for different disorder strength $W \in \{2, 2.5, 3.35, 4.5, 5, 6\}$ with $L$ up to $320$.
The inset shows the data for $W=3.35$ in a log-log scale with a power-law fit (the dashed line) described by $S_{L/2} \propto L^{0.5}$.
(c) $S_{L/2}$ as a function of $W$ for different $L$.
The vertical dashed line denotes the phase boundary between the log-law and area-law regimes at  $W = 3.35$.
(d) The data collapse of the entanglement entropy $S_{L/2}$ using the scaling function Eq.~(\ref{scaling_function}).
Here we use $W \ge 3$ data for scaling collapse.
The inset shows $S_{L/2}$ with respect to $W - W_c$ in a log-log scale, with the solid line being a power-law fit $S_{L/2} \propto (W-W_c)^{-1.09}$.
In (b--d), $\gamma=-0.5$.}
\label{fig1}
\end{figure}

In this work, we study the dynamics of a half-filled many-body state under the evolution of the disordered Hatano-Nelson (HN) model with open boundaries.
We find that the area-law scaling of the entanglement entropy in the disorder-free HN model may develop into a logarithmic scaling at small disorder, that is, $S_{L/2} \propto \log L$.
As we further raise the disorder strength, the system reenters an area-law regime with the entanglement entropy being independent of the system size.
At the critical point between the log-law and area-law regimes, the entanglement entropy exhibits a universal algebraic scaling with $S_{L/2} \sim L^{\beta}$ and $\beta \approx 0.5$.
Based on the entanglement entropy, we map out the phase diagram of the disordered HN model including a log-law and an area-law phase with respect to
the asymmetric hopping strength and the disorder strength [see Fig.~\ref{fig1}(a)].
By examining the subsystem entanglement entropy, the connected correlation function and the mutual information,
we further show that such a log-law phase does not possess a conformal invariance.
Finally, we demonstrate the existence of a phase transition from the log-law to the area-law entanglement for the HN model with periodic boundaries.

\emph{Hatano-Nelson model}.---To study the entanglement phase transition, we consider the paradigmatic HN model with onsite disorder described by the following Hamiltonian~\cite{HN1996PRL}
\begin{equation}
\label{Disordered_HN_Model}
\hat{H} =  \sum_i  \left(J_L\hat{c}_i^\dagger \hat{c}_{i+1} + J_R \hat{c}_{i+1}^\dagger \hat{c}_i +m_i \hat{c}_i^\dagger \hat{c}_i \right),
\end{equation}
where $\hat{c}_i^\dagger$ ($\hat{c}_i$) is the fermionic creation (annihilation) operator at the $i$th site, $J_L=-(J-\gamma)/2$ and $J_R=-(J+\gamma)/2$ with $J$ and $\gamma$
being real parameters depicting the strengths of symmetric and asymmetric hopping, respectively,
and $m_i$ is the on-site disorder uniformly sampled in $[-W/2, W/2]$ with $W$ denoting the disorder strength.
Without disorder ($W=0$), the Hamiltonian under open boundary conditions (OBCs) exhibits non-Hermitian skin effects when $0 < |\gamma| < |J|$, with all single-particle eigenstates localized at one boundary.
In the following, we will set $J=1$ as the units of energy.

To study the entanglement behavior at a sufficiently long time, we consider the following evolving state at time $t$,
\begin{equation}
\label{Evolution}
\ket{\psi (t)} = {e^{-i \hat{H} t} \ket{\psi_0}}/ \lVert{e^{-i \hat{H} t} \ket{\psi_0}}\rVert,
\end{equation}
where $\ket{\psi_0}$ is a many-body state with $L/2$ fermions for a system with $L$ sites ($L$ is even).
We consider the N{\'e}el state $\ket{\psi_0} = \prod_{j=1}^{L/2} \hat{c}_{2j}^\dagger \ket{0}$ as an initial state with $\ket{0}$ denoting the vacuum state.
We note that the dynamics described by Eq.~(\ref{Evolution}) can be realized in an open quantum system by post-selecting a quantum trajectory where no particle loss is detected~\cite{Gullans2021PRL, Ryu2022arXiv}.
Since the Hamiltonian in Eq.~(\ref{Disordered_HN_Model}) is quadratic and the initial state $\ket{\psi_0}$ is a Slater determinant state, the final state $\ket{\psi (t)}$ is also a determinant state and its correlation matrix $D_{ij}(t) = \bra{\psi(t)} \hat{c}_i^\dagger \hat{c}_j \ket{\psi(t)}$ can be efficiently calculated (see Supplemental Material Sec. S-I~\cite{SM} for details).
The von Neumann entanglement entropy $S_A$ between a subsystem $A$ and the rest of the system can be obtained by~\cite{Peschel2003JPA}
\begin{equation}
S_A = - \text{Tr} [D_A  \log D_A + (1 - D_A) \log (1 - D_A)],
\end{equation}
where $D_A$ is the correlation matrix for the subsystem $A$.
In the following, we consider $A = \{ 1,2,...,l \}$ and label $S_A$ as $S_l$.
We note that all the quantities are averaged over $500$ [only for Fig.~\ref{fig1}(a)] or $2000$ random configurations in numerical calculations.

\emph{Entanglement phase transitions under open boundary conditions}.---We now study the entanglement behavior of the state $\ket{\psi(t)}$ at sufficiently long times under OBCs.
For a Hermitian Hamiltonian with $\gamma=0$, the system exhibits a volume-law entanglement when $W=0$, whereas nonzero disorder immediately drives the system to an entanglement area-law phase due to the Anderson localization.
For the non-Hermitian Hamiltonian without disorder, a previous study shows that the entanglement obeys an area law due to the skin effects which push all the particles towards one boundary~\cite{Ryu2022arXiv}.
When disorder is sufficiently strong, we expect that the long-time evolution leads to a state obeying an area law, since 
all the single-particle eigenstates are localized on single sites so that the initial state $\ket{\psi_0}$ stays unchanged.
However, it remains unclear whether there are entanglement phase transitions between these two limiting cases.

\begin{figure}[t]
\includegraphics[width=8.3cm]{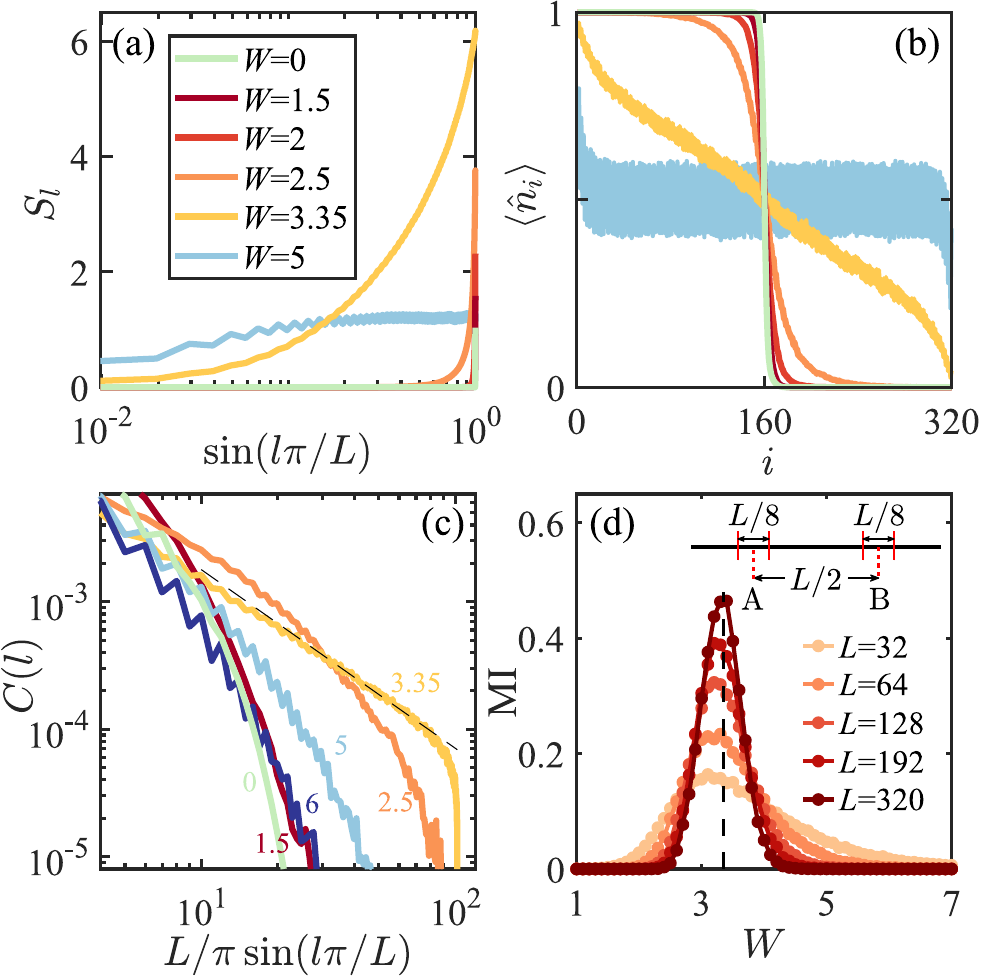}
\caption{(a) The entanglement entropy $S_l$ for subsystems of length $l$ and (b) the density distribution $\langle \hat{n}_i\rangle=\bra{\psi(t)} \hat{c}_i^\dagger \hat{c}_i \ket{\psi(t)} $ of the state $\ket{\psi(t)}$ as a function of site $i$ for different values of disorder strength $W$.
(c) The connected correlation function $C(l)$ with respect to $L/\pi \sin(l\pi/L)$ for $W\in \{ 0, 1.5, 2.5, 3.35, 5, 6 \}$.
The dashed line shows a power-law fit described by $C(l) \propto [L/\pi \sin(l\pi/L)]^{-1.4}$.
In (a--c), the system size $L=320$.
(d) The mutual information $I_{AB}$ with respect to $W$ for different system sizes $L$.
The vertical dashed line marks the phase boundary.
Here, $\gamma=-0.5$.
}
\label{fig2}
\end{figure}

In Fig.~\ref{fig1}(a), we map out the phase diagram based on the entanglement entropy, illustrating the existence of entanglement phase transitions as we increase
the disorder strength $W$. 
In fact, our numerical results (up to $L=320$) suggest that for small disorder strength $W$, the entanglement entropy $S_{L/2}$ of a half system grows logarithmically with the system size $L$, as shown in Fig.~\ref{fig1}(b). 
One can find such a log-law regime in Fig.~\ref{fig1}(a).
The logarithmic growth in this regime can be understood from the fact that weak disorder cannot fully suppress the unidirectional shift induced by the nonreciprocal hopping. 
As a result, the density profile develops a broadened domain wall near the center of the chain, leaving a finite entanglement between the left and right halves [see Fig.~\ref{fig2}(b) and Supplemental Material Sec.~S-II~\cite{SM}].
Further increasing the disorder strength leads to an area-law entanglement, reminiscent of the Anderson localized phase in Hermitian systems.
At the transition point between the log-law and area-law regimes, the entanglement entropy exhibits an algebraic scaling $S_{L/2} \propto L^\beta$ with $\beta \approx 0.5$ [see the inset of Fig.~\ref{fig1}(b)].
These different scaling behaviors can also be clearly observed in Fig.~\ref{fig1}(c), where $S_{L/2}$ grows with $L$ (stays unchanged) for small $W$ (large $W$); the algebraic scaling manifests in a peak around the phase transition point due to a faster growth of entanglement.

To further characterize the entanglement phase transition, we adopt a finite-size scaling form for $S_{L/2}$ given by~\cite{Fisher2018PRB}
\begin{equation}
\label{scaling_function}
S_{L/2}(W,L) = L^\beta F[(W-W_c) L^{1/\nu}],
\end{equation}
with $W \gtrapprox W_c$.
For $W \gg W_c$, one can find that $S_{L/2}$ is independent of the system size $L$, while at $W = W_c$, $S_{L/2}$ scales algebraically with $L$.
Therefore $F(x)$ satisfies
\begin{equation}
F(x)\propto
\left\{
\begin{aligned}
\text{const}&,\ \ \  x=0
\\x^{-\nu \beta}&,\ \ \   x \rightarrow +\infty
\end{aligned}
\right. ,
\end{equation}
leading to $S_{L/2}(W_c,L) \propto L^{\beta}$ at $W=W_c$ and $S_{L/2}(W,L) \propto (W-W_c)^{-\nu \beta}$ for $W \ge W_c$.
By collapsing the $S_{L/2}$ data using the scaling function Eq.~(\ref{scaling_function}), we obtain $W_c = 3.35 \pm 0.05$, $\beta = 0.52 \pm 0.03$ and $\nu = 1.89 \pm 0.05$; the uncertainty corresponds to the standard error of scaling results for different sets of system sizes [see Supplemental Material Sec. S-III~\cite{SM} for details].
The exponents $\beta$ and $\nu$ agree well with those obtained by a direct fit as displayed in the insets of Fig.~\ref{fig1}(b) and (d), where $\beta = 0.5$ and $\beta \nu = 1.09$, respectively.
We also plot the scaled entanglement entropy $S_{L/2} / L^\beta$ as a function of $(W-W_c)L^{1/\nu}$ for $\gamma=-0.5$ in Fig.~\ref{fig1}(d), showing that all the data collapse to a single curve with high quality.
One can also find scaling collapses for other $\gamma$ with similar exponents $\beta \approx 0.5$ and $\nu \approx 1.9$ in Supplemental Material Sec. S-III~\cite{SM}.

Based on the scaling function Eq.~(\ref{scaling_function}), we calculate the transition points for distinct $\gamma$ and mark them out
as the phase boundary in Fig.~\ref{fig1}(a). The boundary corresponds to a large entanglement entropy due to an algebraic scaling of $S_{L/2}$ [see the region marked with bright colors in Fig.~\ref{fig1}(a) and also the peak in Fig.~\ref{fig1}(c)].

One may attribute the entanglement phase transition to the transition of single-particle eigenstates of the HN model from skin modes to Anderson localized states.
In fact, the interplay of non-reciprocal hopping and disorder can result in partially extended single-particle eigenstates for the Hamiltonian where
$\hat{H}=\sum_{i,j} [H]_{ij}\hat{c}_i^\dagger \hat{c}_j$~\cite{Ryu2021PRL, Chen2019PRB}.
To explain such a behavior, we transform $H$ to a Hermitian Hamiltonian ${H}^\prime={H}(J\rightarrow J^\prime, \gamma \rightarrow 0)$ by ${H}^\prime = S^{-1} {H} S$, where $S=\mathrm{diag}\left\{r^{1/2},r,\dots,r^{L/2} \right\}$ with $r = |(J + \gamma)/(J - \gamma)|$ and  $J' = \text{sgn}(J) \sqrt{J^2-\gamma^2}$ for $|\gamma| < |J|$.
The eigenstates of $H'$ are exponentially localized for any non-zero disorder strength $W$, which has an asymptotic form $| u_{n}^\prime (x) |^2\sim e^{-|x - x_n|/\xi} $ with $x_n$ being the localized position and $\xi$ being the localization length.
Based on the similar transformation, we obtain the density profile of the right eigenstates of $H'$, which is given by $ | u_n^{\text{R}}(x)  |^2 \sim r^{x} e^{-|x - x_n|/\xi} $.
Interestingly, for $r e^{1/\xi} = 1$, the density quickly damps to zero as $x$ increases for $x > x_n$ while remains the same for $x < x_n$, so that the state is partially extended in the $x < x_n$ region.
Such a behavior may account for the algebraic scaling of the entanglement entropy at the critical point, which is faster than a log-law but slower than a linear scaling.

However, since the state subject to non-Hermitian evolution is a half-filled many-body state, the entanglement transition point with $W_c\approx 3.35$ clearly deviates from the value of $W_c=3.56$ for the single-particle transition at zero energy in Ref.~\cite{Ryu2021PRL}.
In Supplemental Material Sec. S-IV~\cite{SM}, we use all the single-particle eigenstates to calculate the orthogonality index and the mean inverse participation ratio (MIPR)
and find that the entanglement transition point is very close to the transition point of the orthogonality index and the minimum of MIPR.

\emph{Absence of conformal invariance}.---The logarithmic scaling of the entanglement entropy usually appears in one-dimensional conformal invariant quantum systems, including gapless (critical) Hermitian systems~\cite{Kitaev2003PRL, Cardy2004JSM} and open systems subject to continuous monitoring~\cite{Diehl2021PRL, Saito2022PRL, Pal2022arXiv}.
In the following, we will show that the log-law regime in the disordered HN model under OBCs cannot be characterized by conformal field theories (CFTs).

We first display the subsystem entanglement entropy $S_l$ in Fig.~\ref{fig2}(a).
There, we see that in the log-law regime ($W<W_c$), $S_l$ is zero for most subsystem sizes $l$ while only exhibits a sharp non-zero peak around the center of the system [$\sin(l\pi/L)=1$].
This shows a clear difference with the entanglement entropy predicted by CFTs, which is given by $S_l = (c/6) \log [\sin(\pi l/L)] + s_0$ for a finite system with open boundaries~\cite{Cardy2009JPA}.
Such a behavior is attributable to the non-reciprocal hopping which causes a domain-wall structure in the density profile $\langle \hat{n}_i\rangle$ of $\ket{\psi(t)}$ as displayed in Fig.~\ref{fig2}(b).
If the subsystem $A$ (or the rest of the system) lies in the region where $\langle \hat{n}_i\rangle \approx 0$ or $1$, then we have $S_A \approx 0$ due to the fact that $\rho_A$ is approximately a pure state.
The filled or empty region shrinks as $W$ increases and finally disappears at the critical point, leading to a nonzero $S_l$ for any $l$ at $W=W_c$.
When $W>W_c$, $S_l$ exhibits an area-law behavior [see the cyan line in Fig.~\ref{fig2}(a)].

In addition, we perform numerical calculations of the connected density-density correlation function defined as
\begin{equation}
C(l) = \langle \hat{n}_{L/2} \rangle \langle \hat{n}_{L/2 + l} \rangle - \langle \hat{n}_{L/2} \hat{n}_{L/2 + l} \rangle.
\end{equation}
For the determinant state $\ket{\psi(t)}$, we have $C(l) = | \langle \hat{c}_{L/2}^\dagger \hat{c}_{L/2 + l} \rangle |^2$.
Figure~\ref{fig2}(c) shows that the correlation $C(l)$ exhibits an exponential decay for all $W$ except at the critical point,
in stark contrast to that of conformal invariant systems where $C(l) \sim l^{-2}$~\cite{Peschel2004JSM, Lucas2020PRR, Diehl2021PRL, Pal2022arXiv}.
At the critical point $W=W_c$, $C(l)$ decays algebraically as $C(l) \sim l^{-1.4}$.

We further calculate the mutual information $I_{AB}= S_A + S_B - S_{A\cup B}$ between two disjointed subsystems $A$ and $B$,
which can serve as another indicator for conformal symmetry~\cite{Fisher2019PRB, Diehl2021PRL, Yao2022PRL}.
Figure~\ref{fig2}(d) shows that
in both the log-law and area-law regimes, $I_{AB}$ approaches zero as $L$ increases, in contrast to
a conformal invariant case where the mutual information features a nonzero and constant value for fixed subsystems $A$ and $B$ [e.g., see the inset of Fig.~\ref{fig2}(d)]~\cite{Cardy2009JPA}.
In the log-law regime, this may be caused by the fact that both $A$ and $B$ lies in the region where $\langle \hat{n}_i\rangle = 0$ or $1$.
Intriguingly, we also find that the mutual information grows with $L$ at the critical point.

\begin{figure}[t]
\centering
\includegraphics[width=8.3cm]{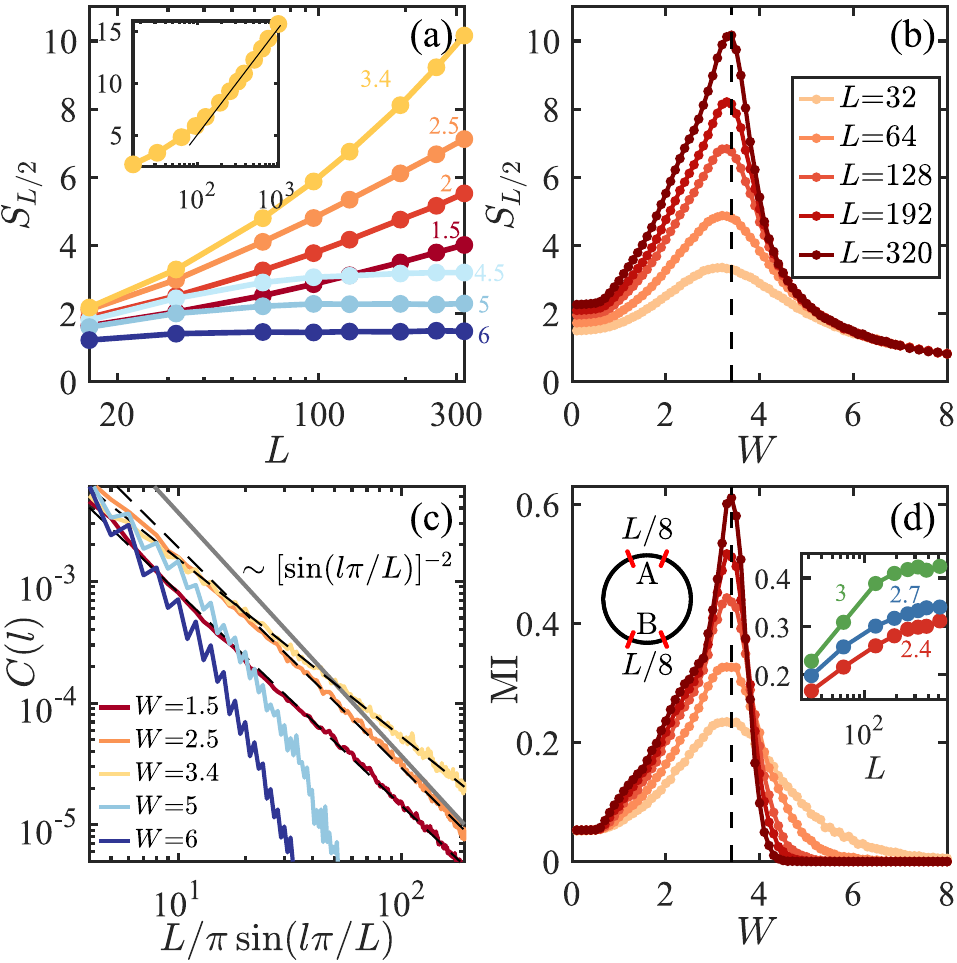}
\caption{(a) The linear-log plot of the entanglement entropy $S_{L/2}$ versus the system size $L$ for $W\in \{ 1.5, 2, 2.5, 3.4, 4.5, 5, 6 \}$.
The inset shows $S_{L/2}$ for $W=3.4$ with $L$ up to $1024$, where the black solid line is a guide for eyes.
(b) The entanglement entropy $S_{L/2}$ as a function of the disorder strength $W$ for various $L$.
(c) The connected correlation function $C(l)$ with respect to $L/\pi \sin(l\pi/L)$ for systems with $L=600$ .
The dashed lines are power-law fits $C(l) \propto [L/\pi \sin(l\pi/L)]^{-\alpha}$ with $\alpha = 1.75$, $1.81$ and $1.46$ for $W=1.5$, $2.5$ and $3.4$, respectively.
We also plot a grey line corresponding to $\alpha=2$ for a visual guide.
(d) The mutual information $I_{AB}$ versus the disorder strength $W$.
The inset displays the $I_{AB}$ as a function of the system size.
The vertical dashed lines in (b) and (d) correspond to $W=3.4$.
}
\label{fig3}
\end{figure}

\emph{Entanglement phase transitions under periodic boundary conditions}.---Next, we study the entanglement properties of the HN model under
periodic boundary conditions (PBCs).
Without disorder, we have proved that
the entanglement entropy of $\ket{\psi(t)}$ at long times scales logarithmically with the system size
as $S_{L/2} = (1/3) \log L$ (see Supplemental Material Sec. S-V~\cite{SM} for the proof).
As the disorder strength $W$ increases, the logarithmic entanglement growth will be destroyed due to the Anderson localization, giving rise to a similar log-to-area-law entanglement phase transition [see Fig.~\ref{fig3}(a) and (b)].
However, in stark contrast to the OBC case where $S_{L/2}$ scales algebraically at the critical point, we find that the entanglement entropy around the critical point tends to converge to a logarithmic scaling under PBCs [see the inset of Fig.~\ref{fig3}(a)].

We further calculate the connected correlation function $C(l)$ under PBCs.
Figure~\ref{fig3}(c) illustrates that in the full log-law regime, the correlation $C(l)$ exhibits an algebraic scaling with $L/\pi \sin(l\pi/L)$, that is, $C(l) \propto [L/\pi \sin(l\pi/L)]^{-\alpha}$.
The power-law fits in Fig.~\ref{fig3}(c) suggest that the exponent $\alpha < 2$ for finite $W$, in contrast to the case for $W=0$ where $\alpha = 2$ (see Supplemental Material Sec. S-V~\cite{SM} for derivation).
For larger $W$, $C(l)$ displays an exponential decay in agreement with the area-law behavior of the entanglement entropy.
To further diagnose the conformal invariance, we calculate the mutual information $I_{AB}$.
Figure~\ref{fig3}(d) shows that in the log-law regime, the mutual information tends to approach constant values for large systems.
However, at the critical point, it exhibits a clear increase as we increase the system size, and
in the area-law regime, it decreases to zero, similar to the OBC case.
Although the system exhibits conformal-like signatures, 
a further numerical study reveals that the logarithmic growth of $S_{L/2}$ in the time direction may be different from that in the spatial direction (see Supplemental Material Sec. S-VIII~\cite{SM}), suggesting that full conformal invariance is limited to the vicinity of $W=0$.

In summary, we have studied the dynamics of a half-filled many-body state of free fermions in the HN model
and found the existence of entanglement phase transitions under both OBCs and PBCs.
For OBCs, although the entanglement entropy obeys the area-law scaling in the case without disorder, our numerical results
suggest that the area-law scaling may develop into a logarithmic scaling in the presence of small disorder.
We further show that the entanglement entropy undergoes a phase transition into an area-law when the disorder becomes sufficiently large.
At the critical point, the entanglement features an algebraic scaling $S_{L/2} \sim L^{\beta}$ with $\beta \approx 0.5$.
Our numerical results also show that the log-law regime under PBCs exhibits conformal-like signatures, while such signatures are absent under OBCs.
We also want to remark that our results are not restricted to the HN model and can be found in other non-Hermitian systems with skin effects, such as the non-Hermitian Su-Schrieffer-Heeger model~\cite{TonyLee, Yao2018PRL1}.
While postselection is required to achieve the non-Hermitian evolution in Eq.~(\ref{Evolution}), it has recently been shown that measurements can induce skin effects
in monitored systems, where the trajectory averaged entanglement entropy obeys an area-law scaling~\cite{ChenFang2022arXiv, ShuChen2022arXiv}.
We thus expect that disorder can drive a similar entanglement phase transition in these systems where postselection is not required.
Our results demonstrate that the interplay between non-Hermitian skin effects and disorder leads to a novel class of entanglement phase transitions,
thereby opening an avenue for exploring entanglement phase transitions in disordered non-Hermitian systems with skin effects.

\begin{acknowledgments}
We thank X. Li and A. Pal for helpful discussions. This work is supported by the National Natural Science Foundation of China (Grant No. 11974201), the Innovation Program for Quantum Science and Technology (Grant No. 2021ZD0301604)
and Tsinghua University Dushi Program.
\end{acknowledgments}

\section*{Author contributions}
K.L. and Z.-C.L. developed the code and performed the numerical and analytical analyses. Y.X. coordinated the work. All authors contributed to the discussion of the results and to the writing of the manuscript.

\clearpage 
\onecolumn
\appendix
\begin{center}
\textbf{\large Supplemental Material: Disorder-Induced Entanglement Phase Transitions in Non-Hermitian Systems with Skin Effects}
\end{center}

\setcounter{equation}{0}
\setcounter{figure}{0}
\setcounter{table}{0}
\setcounter{page}{1}
\renewcommand{\theequation}{S\arabic{equation}}
\renewcommand{\thefigure}{S\arabic{figure}}
\renewcommand{\thetable}{S\arabic{table}}
\renewcommand{\thesection}{S-\Roman{section}}

In the Supplemental Material, we will elaborate on how to calculate the entanglement entropy and justify the choice of time step in Section S-I,
present the physical intuition for the log-law regime under OBC in Section S-II,
provide more details about the scaling collapse of the entanglement entropy in Section S-III,
present the orthogonality index and mean inverse participation ratio in Section S-IV,
prove that the long-time entanglement entropy for the pristine HN model under PBCs obeys a logarithmic scaling in Section S-V,
clarify the limitations of determining the steady state from the single-particle eigenstates in Section S-VI,
discuss the relation between the single-particle spectrum and the entanglement under PBCs in Section S-VII,
and examine the spacetime conformal symmetry under PBCs in Section S-VIII.

\section{Details on how to calculate the entanglement entropy}
In this section, we will elaborate on how to calculate the dynamics of the entanglement entropy (also see Ref.~\cite{Ryu2022arXiv}).
We consider an initial determinant state $\ket{\psi_0}$ evolved by a non-Hermitian free fermion Hamiltonian $\hat{H} = \sum_{ij} H_{ij} \hat{c}_i^\dagger \hat{c}_j$.
The dynamics is governed by
\begin{equation}
	\ket{\psi (t)} = \frac{e^{-i \hat{H} t} \ket{\psi_0}}{\sqrt{\bra{\psi_0} e^{i \hat{H}^\dagger t} e^{-i \hat{H} t} \ket{\psi_0} } }.
\end{equation}
Without loss of generality, we choose the N{\'e}el state as an initial state, that is, $\ket{\psi_0} = \prod_{i=1}^{L/2} \hat{c}_{2i}^\dagger \ket{0}$.
Then, the evolving state $\ket{\psi(t)}$ at time $t$ can be written as
\begin{equation}
	\begin{aligned}
		\ket{\psi (t)} &= \frac{1}{\sqrt{N(t)} }  e^{-i \hat{H} t} \prod_{i=1}^{L/2} \hat{c}_{2i}^\dagger \ket{0}
		= \frac{1}{\sqrt{N(t)} } \prod_{i=1}^{L/2} \hat{c}_{2i}^\dagger(t) e^{-i \hat{H} t} \ket{0}
		\\&= \frac{1}{\sqrt{N(t)} } \prod_{i=1}^{L/2} \hat{c}_{2i}^\dagger(t) \ket{0},
	\end{aligned}
\end{equation}
where $N(t) = \bra{\psi_0} e^{i \hat{H}^\dagger t} e^{-i \hat{H} t} \ket{\psi_0}$ and $\hat{c}_{i}^\dagger (t) = e^{-i \hat{H} t} \hat{c}_{i}^\dagger e^{i \hat{H} t} = \sum_{j=1}^{L} [e^{-iHt}]_{ji} \hat{c}_j^\dagger$.
We see that $\ket{\psi(t)}$ is also a determinant state except that the operators $\hat{c}_{2i}^\dagger (t)$ are not necessarily orthogonal.
We can write the unnormalized evolving state as
\begin{equation} \label{UnEvoState}
	\ket{\tilde{\psi}(t)} = \prod_{i=1}^{L/2} \hat{c}_{2i}^\dagger(t) \ket{0}
	= \left[ \prod\nolimits_{j=1}^{L/2} \left( \sum\nolimits_{i=1}^L [U(t)]_{ij} \hat{c}_i^\dagger \right) \right] \ket{0},
\end{equation}
where
\begin{equation}
	U(t) = e^{-i H t} U_0
\end{equation}
with $U_0$ (an $L\times \frac{L}{2}$ matrix) being a collection of all the initial single-particle states with $[U_0]_{ij} = \delta_{i,2j}$.

We now apply a QR decomposition on $U(t)$ and obtain $U(t) = QR$, where $Q$ is an $L\times \frac{L}{2}$ matrix satisfying $Q^\dagger Q = 1$ and $R$ is an upper-triangular matrix. Substituting the decomposition into Eq.~(\ref{UnEvoState}) yields
\begin{equation}
	\ket{\tilde{\psi}(t)} = \left[ \prod\nolimits_{j=1}^{L/2} \left(  \sum\nolimits_{k=1}^{L/2} R_{kj} \sum\nolimits_{i=1}^L  Q_{ik} \hat{c}_i^\dagger \right) \right] \ket{0}.
\end{equation}
Let us define $\hat{\gamma}_k^\dagger = \sum_{i=1}^L Q_{ik} \hat{c}_i^\dagger$. They are fermionic creation operators because  they satisfy
the anti-commutation relations, i.e., $\{ \gamma_i^\dagger, \gamma_j^\dagger \} = 0$ and $\{ \gamma_i, \gamma_j^\dagger \} = \delta_{ij}$.
It follows that the unnormalized evolving states can be written in terms of these new operators as
\begin{equation}
	\begin{aligned}
		\ket{\tilde{\psi}(t)} &= \left[ \prod\nolimits_{j=1}^{L/2} \left(  \sum\nolimits_{k=1}^{L/2} R_{kj} \gamma_k^\dagger \right) \right] \ket{0}
		\\&= (R_{1,\frac{L}{2}} \gamma_1^\dagger + \cdots + R_{\frac{L}{2},\frac{L}{2}} \gamma_{\frac{L}{2}}^\dagger) \cdots (R_{13} \gamma_1^\dagger + R_{23} \gamma_2^\dagger + R_{33} \gamma_3^\dagger) (R_{12} \gamma_1^\dagger + R_{22} \gamma_2^\dagger) (R_{11} \gamma_1^\dagger) \ket{0}
		\\&= (R_{\frac{L}{2},\frac{L}{2}} \gamma_{\frac{L}{2}}^\dagger) \cdots (R_{33} \gamma_3^\dagger) (R_{22} \gamma_2^\dagger) (R_{11} \gamma_1^\dagger)  \ket{0}
		\\&= \left( \prod\nolimits_{i=1}^{L/2} R_{ii} \gamma_i^\dagger \right) \ket{0}.
	\end{aligned}
\end{equation}
Since $\langle \tilde{\psi}(t) | \tilde{\psi}(t) \rangle = (\prod_{i=1}^{L/2} R_{ii})^2$, we arrive at
\begin{equation}
	\ket{\psi(t)} = \prod\nolimits_{i=1}^{L/2} \gamma_i^\dagger \ket{0}.
\end{equation}
The correlation function of the final state, defined as $D_{ij}(t) = \bra{\psi(t)} \hat{c}_i^\dagger \hat{c}_j \ket{\psi(t)}$, can be easily calculated by
\begin{equation}
	D(t) = (Q Q^\dagger)^{T}.
\end{equation}
We then can evaluate the von Neumann entanglement entropy $S_A$ between a subsystem $A$ and the rest of the system by~\cite{Peschel2003JPA}
\begin{equation}
	S_A = - \text{Tr} [D_A  \log D_A + (1 - D_A) \log (1 - D_A)],
\end{equation}
where $D_A$ denotes the correlation matrix for the subsystem $A$.

In principle, one can perform a single QR decomposition to obtain the correlation function $D(t)$ even if $t$ is large.
However, since the Hamiltonian is non-Hermitian, the elements in $U(t)$ may grow or decay exponentially with $t$.
To avoid numerical instabilities, we perform a QR decomposition for every time step $\Delta t$, i.e.,
\begin{equation}
	U(t + \Delta t) = \text{qr}[e^{-i H \Delta t} U(t)],
\end{equation}
where qr stands for the QR decomposition.

\begin{figure}
	\centering
	\includegraphics[width=0.35\textwidth]{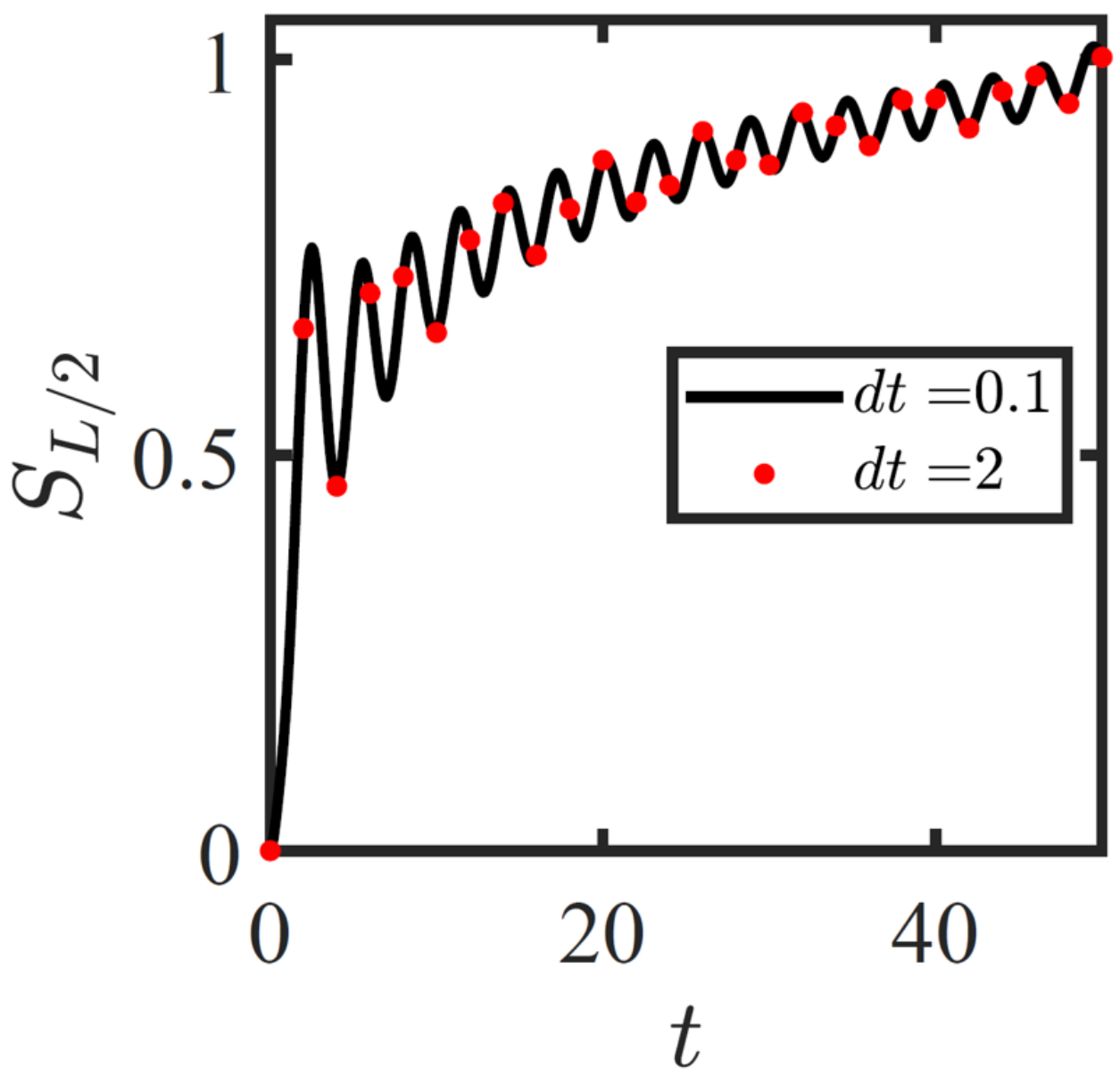}
	\caption{The entanglement entropy $S_{L/2}$ as a function of the evolution time $t$ for different choices of $\Delta t$.}
	\label{fig_different_dt}
\end{figure}

In our numerical calculations, we set $\Delta t = 2$ and $N_t = 1000$ ($N_t$ denotes the number of time steps).
Since the QR decomposition, which is used to keep the numerical calculation stable, does not change the underlying determinant state, the results are independent of the choice of $\Delta t$ for a fixed evolution time $t=N_t \Delta t$.
We have checked this by comparing the entanglement entropy obtained with different $\Delta t$ in Fig.~\ref{fig_different_dt}.

For all the quantities in the main text, we average over the last $100$ time steps as well as $500$ or $2000$ disorder realizations.
The time evolution is numerically simulated by using the matrix exponential function ``expm'' in a MATLAB program.
In the simulation, we use $[U_0]_{ij} = \delta_{i,2j}$ and $\gamma <0$.
Here we note that for $\gamma > 0$ the numerical results for OBCs may be incorrect when the system size is large.
The error might arise from the skin effect which tends to make the columns of $U(t)$ similar to each other and hard to be orthogonalized.
We have checked the correctness of our results up to $L=320$ by increasing the numerical precision using MATLAB's \texttt{vpa} function.

\section{Physical intuition for the log-law regime under OBC}

\begin{figure}
	\centering
	\includegraphics[width=\textwidth]{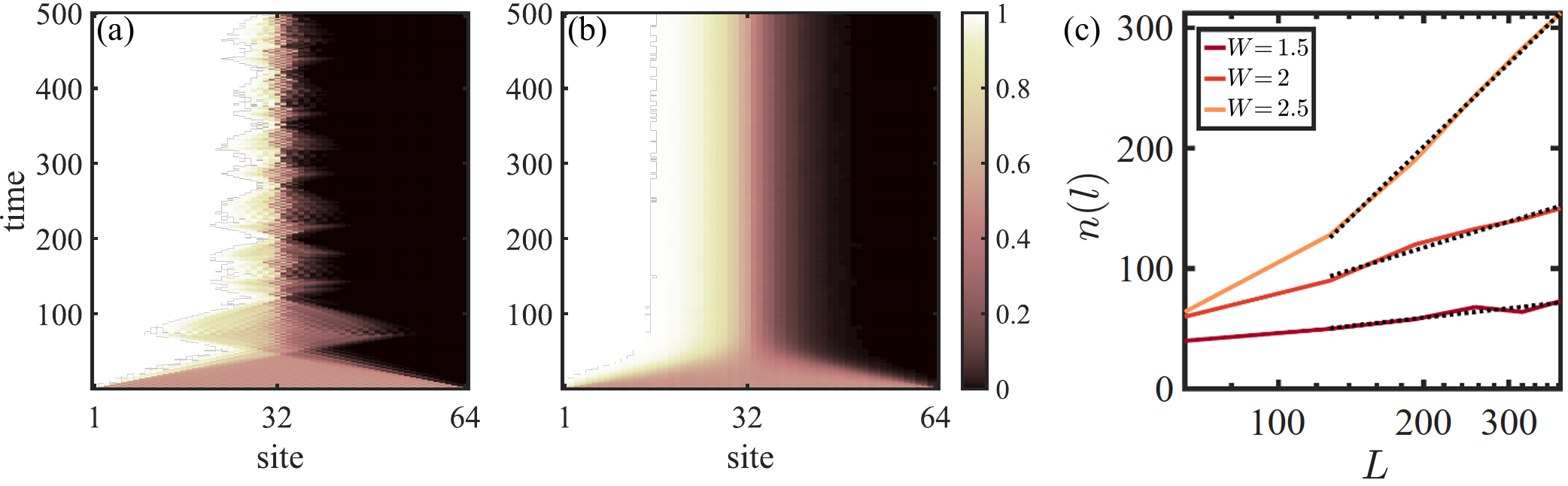}
	\caption{The density distribution $\langle \hat{n}_i\rangle$ as a function of site $i$ and time $t$ under OBCs for (a) $W=0$ and (b) $W=2$, with system size $L=64$.
	(c) The domain-wall size as a function of the system size $L$ for different disorder strengths $W$, where the dashed lines are logarithmic fits.
	Here we set $J=1$, and $\gamma=-0.5$.
	}
	\label{fig_n_vs_t}
\end{figure}

In a non-Hermitian system without disorder, 
the asymmetric hopping term causes particles to move unidirectionally, 
resulting in a nearly perfect product state 
$\ket{1 1 \cdots 1 0 0 \cdots 0}$ with a sharp domain wall 
[see Fig.~\ref{fig_n_vs_t}(a)]. 
Introducing small disorder cannot cause the particles to freeze as in a Hermitian system, 
but allows particles to scatter during their transfer to the boundary. 
This scattering results in an approximate product state with a smooth transition of particle number around the middle of the chain 
[see Fig.~\ref{fig_n_vs_t}(b)].

Based on the definition of the entanglement entropy, 
we know that the bipartite entanglement entropy of a state $\ket{\psi}$ is equal to that of the state 
$\ket{1}^{\otimes n} \otimes \ket{\psi} \otimes \ket{0}^{\otimes n}$ 
given that the two partitions $A$ and $B$ are the left and right halves of the system, respectively. 
As such, only the central part with particle numbers between $0$ and $1$ contributes to the entanglement entropy. 
We identify the sites supporting $\ket{\psi}$ as the domain wall, spanning from light yellow to dark pink in Fig.~\ref{fig_n_vs_t}(a--b).

Numerically, we define the size of the domain wall as the range where the average particle number lies between $10^{-5}$ and $1 - 10^{-5}$. 
Our calculations show that the domain wall size scales logarithmically with the system size [see Fig.~\ref{fig_n_vs_t}(c)], which may contribute to the logarithmic growth of the bipartite entanglement entropy of the system.

Notably, this behavior appears specific to random disorder. 
In the case of quasiperiodic disorder, the domain wall size remains independent of the system size, leading to area-law entanglement instead. 
We conjecture that random disorder induces a stronger localization effect, which can counteract the unidirectional current and result in a broadened domain wall. 
In contrast, quasiperiodic disorder typically has a weaker localization effect which cannot suppress the unidirectional particle flow, 
leading to perfect particle accumulation at the left half of the system and an area-law entanglement.

\section{Details on the scaling collapse of the entanglement entropy}
\begin{figure}
	\centering
	\includegraphics[width=\textwidth]{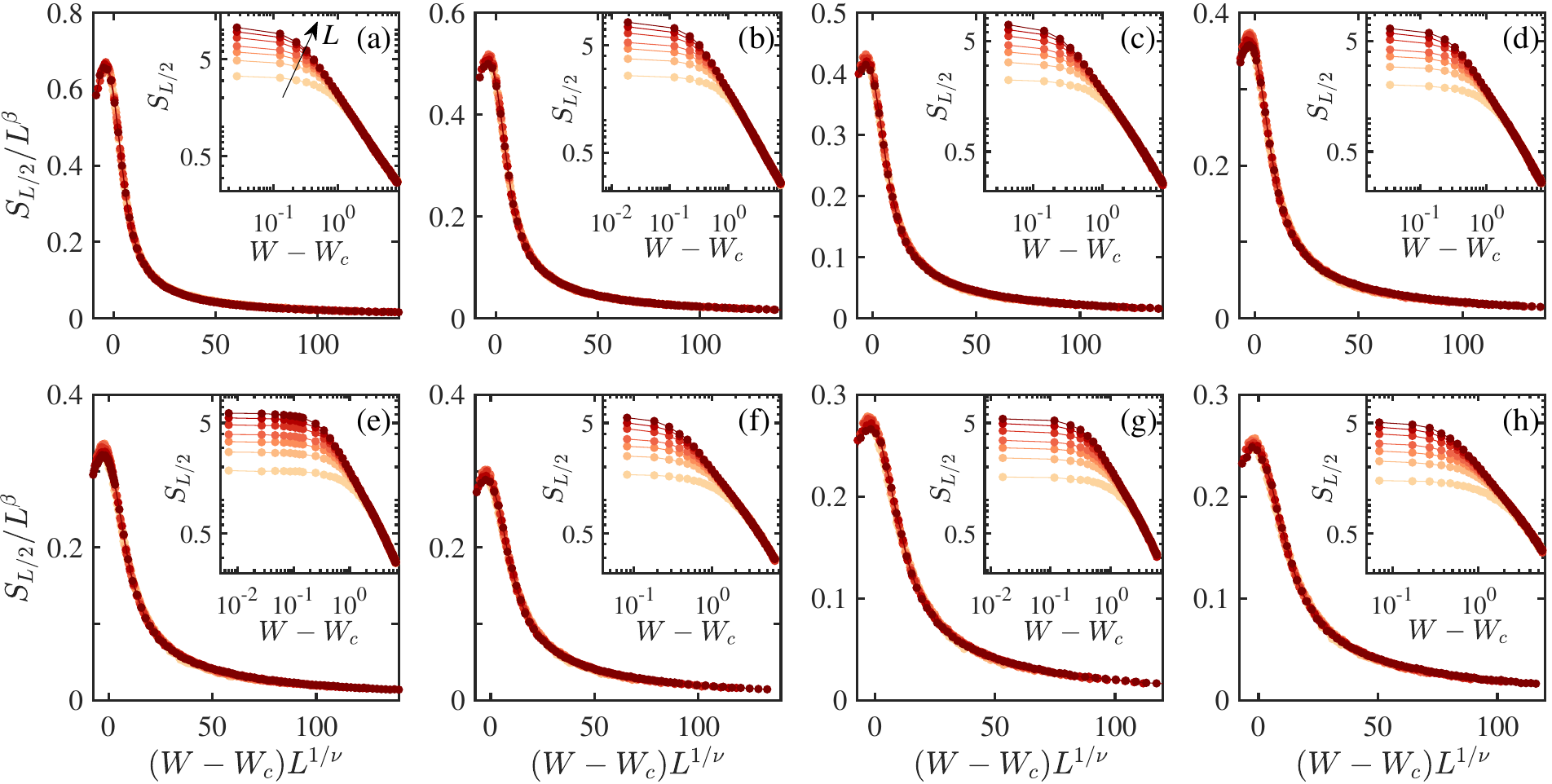}
	\caption{(a--h) The scaling collapses of the entanglement entropy for $\gamma = -0.1, -0.2, -0.3, -0.4, -0.5, -0.6, -0.7, -0.8$.
		Here the system size $L \in \{ 32, 64, 96, 128, 192, 256, 320 \}$ and we use brighter (deeper) colors to denote smaller (larger) system sizes.
		The insets show $S_{L/2}$ versus $W-W_c$ in a log-log scale.
	}
	\label{figS1}
\end{figure}

In this section, we will provide more details about the scaling collapse of the entanglement entropy.
We use Eq.~(4) in the main text to perform finite-size scaling, which can be rewritten as
\begin{equation}
	S_{L/2}(W,L)/L^\beta =  F[(W-W_c) L^{1/\nu}].
\end{equation}
Let us define $y(x,L) = S_{L/2}(W,L)/L^\beta$ where $x=(W-W_c) L^{1/\nu}$.
One needs to find an optimal set of parameters $\{ W_c, \nu, \beta \}$ such that $y(x,L)$ versus $x$ lines for different $L$ collapse to a single curve.
This can be done by minimizing the loss function defined as
\begin{equation}
	\label{lossfunc}
	\mathcal{L} = \sum_{x,L}  [ y(x,L) - \bar{y} (x) ]^2,
\end{equation}
where $\bar{y}(x) = \sum_L  y(x,L)/N_L$ with $N_L$ denoting the number of $L$ over the sum.
We use the \texttt{fminsearch} function in a MATLAB program to find the optimal parameters that minimize $\mathcal{L}$.
To estimate the uncertainty of the parameters, we extract the parameters for different sets of system sizes and evaluate the standard deviation.
The data collapse and the extracted parameters are shown in Fig.~\ref{figS1} and Table~\ref{tableS1}, respectively.

\begingroup

\setlength{\tabcolsep}{3pt} 
\renewcommand{\arraystretch}{1.25} 

\begin{table}[b]
	\caption{Parameters extracted by minimizing the loss function Eq.~(\ref{lossfunc}) for different $\gamma$.
	}
	\begin{tabular}{c c c c c c c c c}
		\hline \hline
		$\gamma$	& $-0.1$			& $-0.2$			& $-0.3$			& $-0.4$			& $-0.5$			& $-0.6$			& $-0.7$			& $-0.8$			\\ \hline
		$W_c$		& $1.47 \pm 0.01$	& $2.08 \pm 0.02$	& $2.56 \pm 0.02$	& $2.97 \pm 0.04$	& $3.35 \pm 0.05$	& $3.72 \pm 0.05$	& $4.08 \pm 0.05$	& $4.43 \pm 0.06$	\\
		$\nu$		& $1.99 \pm 0.04$	& $1.91 \pm 0.05$	& $1.88 \pm 0.05$	& $1.88 \pm 0.06$	& $1.89 \pm 0.05$	& $1.89 \pm 0.04$	& $1.89 \pm 0.03$	& $1.90 \pm 0.03$	\\
		$\beta$		& $0.51 \pm 0.01$	& $0.50 \pm 0.02$	& $0.51 \pm 0.02$	& $0.51 \pm 0.03$	& $0.52 \pm 0.03$	& $0.52 \pm 0.02$	& $0.53 \pm 0.02$	& $0.53 \pm 0.02$	\\\hline\hline
	\end{tabular}
	\label{tableS1}
\end{table}

\endgroup

\section{Orthogonality index and mean inverse participation ratio}
In the main text, we have argued that the entanglement phase transition for a half-filled many-body state may be related to the single-particle phase
transition from skin states to Anderson localized states. In this section, we will show that
the entanglement phase transition point is very close to the transition point of the orthogonality index and the minimum of the mean inverse participation ratio (MIPR)
calculated using all the single-particle eigenstates.

Since the skin states are almost linearly dependent, we thus introduce the orthogonality index defined as $O=|\det (U)|^{1/L}$ to characterize the phase transition
from skin states to Anderson localized ones.
Here, $U=(\ket{u_1^{\text{R}} },...,\ket{u_L^{\text{R}} })$ with $\ket{u_n^{\text{R}} }$ ($n=1,\dots,L$) being the normalized right eigenstates of the Hamiltonian.
The index $O$ characterizes the degree of orthogonality for the set of all eigenstates $\{ \ket{u_n^{\text{R}} } \}_{n=1}^L$.
If $O = 1$, the eigenstates are orthogonal; if $O < 1$, they are non-orthogonal.
Specifically, $O$ approaches zero if the set of eigenstates are almost linearly dependent, which is the case for a set of skin states.

\begin{figure}
	\centering
	\includegraphics[width=9.5cm]{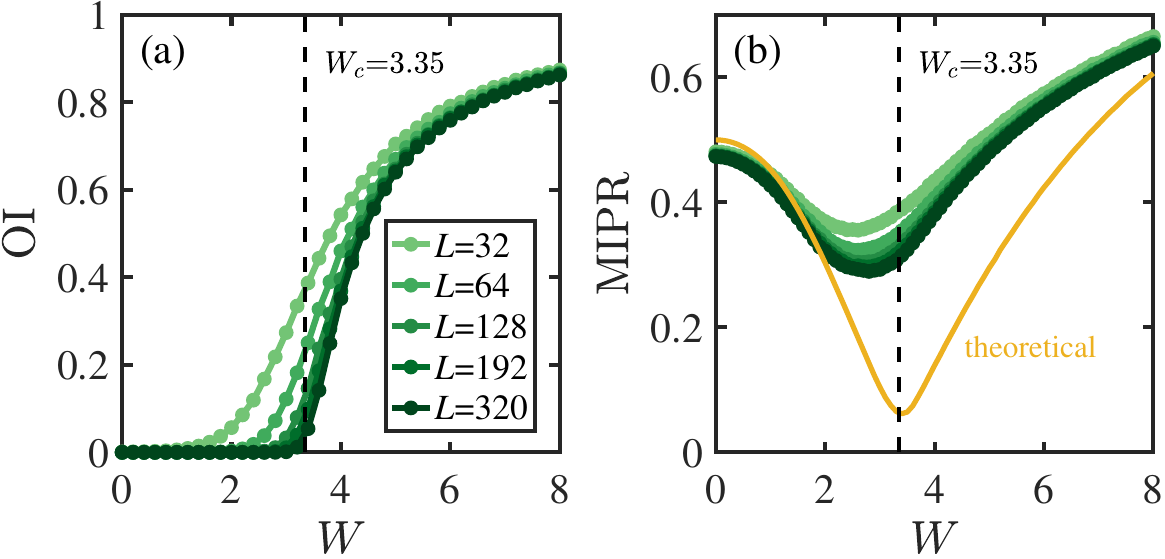}
	\caption{(a) The orthogonality index $O$ and (b) the MIPR as a function of disorder strength $W$ for various system sizes.
		The vertical dashed lines in (a) and (b) mark the entanglement phase transition point $W_c = 3.35$.
		Here, $\gamma = -0.5$.
	}
	\label{figS2}
\end{figure}

Figure~\ref{figS2}(a) displays the orthogonality index $O$ for $\gamma = -0.5$.
We see the existence of two phases: one with vanishingly small values of $O$ and the other with finite values.
The phase transition from skin states to Anderson localized states is revealed by a sharp rise of $O$ from nearly zero to non-zero values for large $L$.
Although we cannot identify the transition point exactly due to finite-size effects, the results indicate that the transition point is very close to the
entanglement phase transition point at $W_c = 3.35$.

We further employ the MIPR to characterize the phase transition.
The MIPR is defined as $I_{\text{M}} = (\sum_n I_n)/L$ where $I_n=\sum_{x} | u_n^{\text{R}}(x)  |^4$ is the inverse participation ratio (IPR) for a normalized right eigenstate $u_n^{\text{R}}(x)$.
When all the states are extended, the MIPR is small, approaching zero as the system size increases, whereas when all the states are localized at one site, $I_m=1$.
Fig.~\ref{figS2}(b) illustrates that the MIPR for the disordered HN model is large when $W$ is either small or large,
consistent with the fact that both skin states and Anderson localized states are spatially localized.

Interestingly, the MIPR decays to a minimum for some disorder strength $W$ as shown in Fig.~\ref{figS2}(b)
(the existence of a minimum in the MIPR has also been found in a non-Hermitian quasicrystal~\cite{Chen2019PRB}).
As shown in the main text, the density distribution of $u_{n}^{\text{R}} (x)$ has an asymptotic form given by
\begin{equation}
	\label{asymptotic_form}
	| u_n^{\text{R}}(x) |^2 \sim r^{x} e^{-|x - x_n|/\xi} = \left\{
	\begin{aligned}
		& e^{- x_n/\xi}\, (r  e^{1/\xi})^x ,\ \ \  x < x_n
		\\ & e^{x_n/\xi}\, (r e^{-1/\xi})^x ,\ \ \   x > x_n
	\end{aligned}
	\right.
\end{equation}
which will exhibit different behaviors for different values of the localization length $\xi$.
We assume that $\gamma < 0$ so that $r e^{-1/\xi} < 1$, which means that the density for the $x > x_n$ part is always suppressed as $x$ increases.
For $r e^{1/\xi} < 1$, the density will get enhanced as $x$ decreases for $x < x_n$, so that the transformed state is a skin state localized at the left boundary.
While for $r e^{1/\xi} > 1$, the state is exponentially localized at $x_n$, giving rise to an Anderson localized state.
For $r e^{1/\xi} = 1$, as also mentioned in the main text, the density remains the same for $x < x_n$ while quickly damps to zero as $x$ increases for $x > x_n$, rendering the state partially extended in the region $x < x_n$.
Thus, the states become more extended as $W$ approaches the phase transition point, accounting for the fact that the MIPR reaches a minimum around the critical point.

To eliminate the finite-size effects, we further calculate the MIPR based on
\begin{equation}
	I_{\text{M}}=\int dE D(E)I[\xi(E)]/\int dE D(E),
\end{equation}
where $\int dE$ denotes an integral over energy $E$,
$D(E)$ is the density of state of the similar transformed Hamiltonian $H^\prime$,
and $I[\xi(E)]$ is the IPR computed based on the asymptotic form Eq.~(\ref{asymptotic_form}) at energy $E$, with the localization length $\xi(E)$ determined by the transfer matrix method~\cite{Kramer1983ZPB}. Fig.~\ref{figS2}(b) plots the calculated MIPR [see the yellow line in Fig.~\ref{figS2}(b)], showing the existence of a dip around $W=3.4$, which is very close to the critical point $W_c=3.35$ for the entanglement phase transition.

\section{Long-time entanglement entropy for the pristine HN model under PBCs}
In this section, we will show that the long-time entanglement entropy for the HN model under PBCs without disorder obeys a logarithmic scaling.
We first prove that the system will converge to a state with the largest imaginary eigenenergy under the evolution of a non-Hermitian Hamiltonian.

We denote the many-body eigenenergy in the $N$-particle subspace as $E_n = \varepsilon_{n}^{\text{R}} + i \varepsilon_{n}^{\text{I}}$ and the corresponding right (left) eigenstate as $\ket{\phi_n^R}$ ($\bra{\phi_n^L}$).
Given an initial $N$-particle determinant state $\ket{\psi_0}$, the evolving state $\ket{\psi(t)}$ can be written as
\begin{equation}
	\begin{aligned}
		\ket{\psi (t)} &= \frac{\sum_n e^{-i  \varepsilon_{n}^{\text{R}}  t} e^{ \varepsilon_{n}^{\text{I}} t} \ket{\phi_n^R} \langle \phi_n^L | \psi_0 \rangle}{\sqrt{\bra{\psi_0} e^{i \hat{H}^\dagger t} e^{-i \hat{H} t} \ket{\psi_0} }}
		\\ &=  \frac{e^{\varepsilon_{1}^{\text{I}} t}}{\sqrt{\bra{\psi_0} e^{i \hat{H}^\dagger t} e^{-i \hat{H} t} \ket{\psi_0} }} \sum_n
		e^{-i \varepsilon_{n}^{\text{R}}  t} e^{(\varepsilon_{n}^{\text{I}} -\varepsilon_{1}^{\text{I}}) t} \ket{\phi_n^R} \langle \phi_n^L | \psi_0 \rangle.
	\end{aligned}
\end{equation}
By assuming $\varepsilon_{1}^{\text{I}}> \varepsilon_{2}^{\text{I}} > \dots$, one can find that $\ket{\psi(+\infty)} = \ket{\phi_1^R}$ up to a phase factor.
Therefore, the entanglement entropy at long times is the same as that of the many-body eigenstate with the largest imaginary eigenenergy, if only a single
eigenstate has the largest imaginary eigenenergy. Otherwise,
the final state will become a superposition of the eigenstates with the largest imaginary eigenenergy.

For the HN model without disorder under PBCs,
the single-particle eigenstates are given by $\ket{k} = \hat{c}_k^\dagger \ket{0} = \frac{1}{\sqrt{L}} \sum_j e^{ikj} \hat{c}_j^\dagger \ket{0}$ corresponding to eigenenergies
$E_k = - J \cos k + i \gamma \sin k$ where $k = k_n = 2\pi n /L$ with $n=-L/2, -L/2+1, ..., L/2-1$. Without loss of generality, we will consider the case with $\gamma<0$ in the following.
Starting from an initially half-filled state, the system will converge to a superposition of
$\ket{\Psi_1} = \prod_{n=-L/2}^{-1} \hat{c}_{k_n}^\dagger \ket{0}$ and
$\ket{\Psi_2} = \prod_{n=-L/2+1}^{0} \hat{c}_{k_n}^\dagger \ket{0}$, which are many-body eigenstates with the largest imaginary eigenenergy in the half-filled subspace.
For infinitely large $L$, the final state is a Slater determinant of all the Bloch states with momentum $k \in [-\pi, 0]$, whose correlation matrix is given by
\begin{equation}
	D_{mn} = \int_{-\pi}^{0} \frac{dk}{2 \pi}  e^{-ik(m-n)} = \frac{i}{2\pi} \frac{1-e^{i \pi (m-n)}}{m - n}.
\end{equation}
When $m-n$ is even, $D_{mn}=0$; otherwise, $D_{mn}=i/[\pi(m-n)]$ so that $|D_{mn}|^2=1/[\pi^2 (m-n)^2]$, indicating that
the density-density correlation function $C(l)\propto 1/l^2$.
The entanglement entropy is determined by the eigenvalues of $D_A$ with $[D_A]_{mn} = D_{mn}$ for $m,n \in A$.

To evaluate the entanglement entropy given by $D_A$, we consider the ground state of a Hermitian free fermion chain with Hamiltonian
$\hat{H}_h = - \sum_j (\hat{c}_j^\dagger \hat{c}_{j+1} + \text{H.c.})$.
Since the ground state of $\hat{H}_h$ is a Slater determinant of Bloch states with momentum $k \in [-\pi/2, \pi/2]$, its correlation matrix is given by
\begin{equation}
	D_{mn}^\prime = \int_{-\pi/2}^{\pi/2} \frac{dk}{2 \pi}  e^{-ik(m-n)} = \frac{\sin[\pi (m-n)/2]}{\pi (m-n)},
\end{equation}
which is related to $D$ by a unitary transformation $D^\prime = U^\dagger D U$ with $U = \text{diag}\{i,i^2,i^3,...\}$ (similar property holds for $D_A$ and $D_A^\prime$).
Based on a continuum approximation, it has been proved that the entanglement entropy of the ground state of $\hat{H}_h$ is asymptotically given by $S = \frac{1}{3} \ln L$ at large $L$~\cite{Peschel2004JSM, Peschel2009JPA}.
We thus conclude that the long-time entanglement entropy for the HN model under PBCs is also given by $S = \frac{1}{3} \ln L$ for sufficiently large $L$,
owing to the fact that $D_A$ and $D_A^\prime$ share the same eigenvalues.
We have also numerically checked that the entanglement entropy is actually described by $S = k \ln L + b$ with $k$ being exactly $1/3$ and $b\approx0.34$.
The non-zero intercept $b$ is attributed to the fact that the derivation for the eigenvalues of $D_A'$ requires a continuum approximation~\cite{Peschel2004JSM}, which is not valid for small $L$.

\section{Limitations of determining the steady state from the single-particle eigenstates}

One may expect that the steady state can be obtained by directly diagonalizing the non-Hermitian Hamiltonian.
Here we show that the steady state in the half-filling sector cannot be uniquely determined based on the single-particle spectrum, 
making it difficult to select a single eigenstate as the steady state.

\begin{figure}
	\centering
	\includegraphics[width=0.85\textwidth]{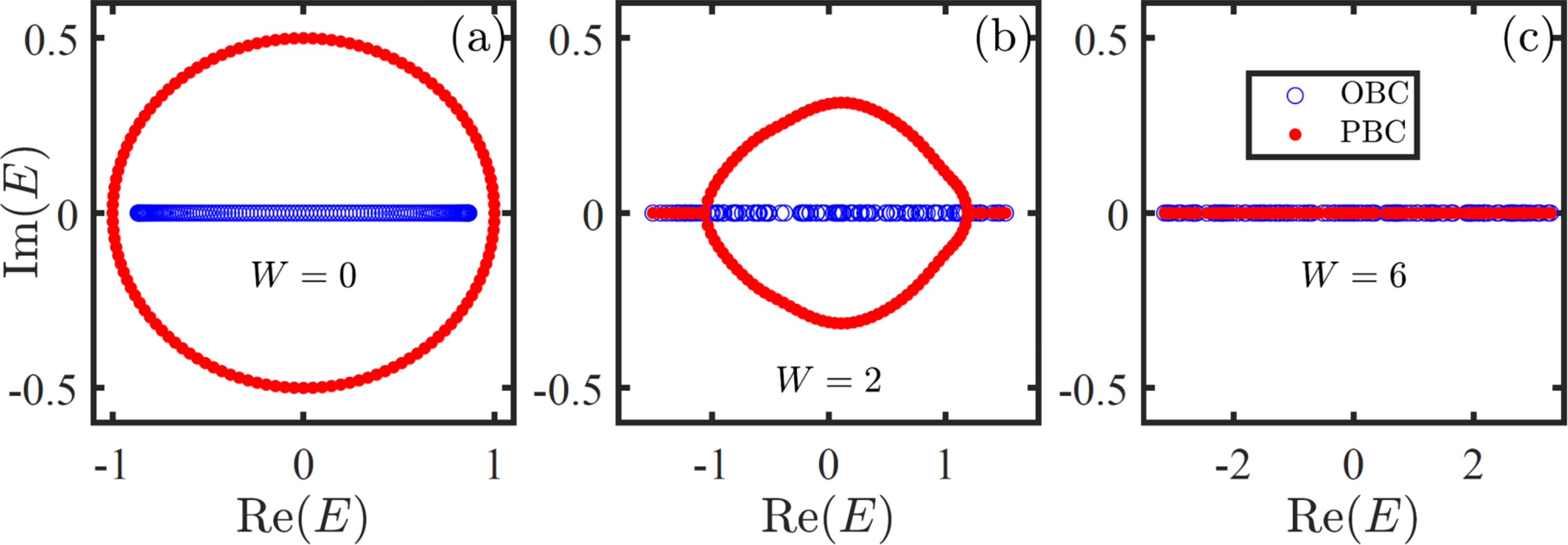}
	\caption{The single-particle spectrum of the disordered HN model under OBCs and PBCs for various $W$.}
	\label{fig_OBC_PBC_spectrum}
\end{figure}

As discussed in Sec.~S-V, the long-time dynamics is governed by the many-body eigenstates with the largest imaginary part of the eigenenergy. 
Due to the quadratic structure of $\hat{H}$, 
an $N$-particle many-body eigenstate can be written as a Slater determinant of $N$ single-particle eigenstates, 
and its many-body eigenenergy is the sum of the $N$ single-particle eigenenergies.
Furthermore, the spectrum of the disordered HN model is symmetric with respect to the real axis, since $H^* = H$.

Under OBCs, the single-particle spectrum is purely real [see Fig.~\ref{fig_OBC_PBC_spectrum}(a)], thus the many-body eigenenergies are also real. 
As all the eigenstates share the same imaginary eigenenergy, one cannot select a single many-body eigenstate as the steady state. 
Instead, the steady state is a superposition of all half-filled eigenstates when a N{\'e}el state is chosen as the initial state.

Under PBCs, our numerical results show that when $W \neq 0$, 
the single-particle spectrum exhibits significant degeneracy at $\mathrm{Im}(E) = 0$ [see Fig.~\ref{fig_OBC_PBC_spectrum}(b,c)]. 
Since the spectrum is symmetric with respect to the real axis, 
the many-body eigenstates with largest imaginary eigenenergies in the half-filling sector should be highly degenerate, 
making it challenging to pick a representative eigenstate as the steady state.

\section{Relation between the single-particle spectrum and the entanglement under PBCs}

When $W\neq 0$ and $W$ is not very large, the single-particle spectrum of $H$ still forms a loop, 
but with part of the spectrum distributed along the real axis outside the loop [see Fig.~\ref{fig_OBC_PBC_spectrum}(b)].
The single-particle states on the loop are delocalized and contribute to the entanglement, while the states on the real axis are localized due to Anderson localization~\cite{HN1996PRL}.
In the case of half-filling, when the system reaches the steady state, all states in the upper half of the loop with $\mathrm{Im}(E)>0$ and half of the states on the real axis will be occupied.
Since the states on the real axis are Anderson localized states and do not contribute to the entanglement, we end up with a logarithmic entanglement scaling similar to the $W=0$ case when $W$ is not very large.

When $W$ is large, the loop structure in the spectrum disappears, and the spectrum lies entirely on the real axis [see Fig.~\ref{fig_OBC_PBC_spectrum}(c)].
Consequently, all the single-particle states are Anderson localized, resulting in an area-law scaling of the entanglement entropy.
Therefore, the transition of the entanglement entropy from logarithmic to area-law scaling with the disorder strength $W$ in PBCs can be captured by the disappearance of the loop structure in the single-particle spectrum of $H$.

\section{Spacetime conformal symmetry under PBCs}

\begin{figure}
\centering
\includegraphics[width=0.6\textwidth]{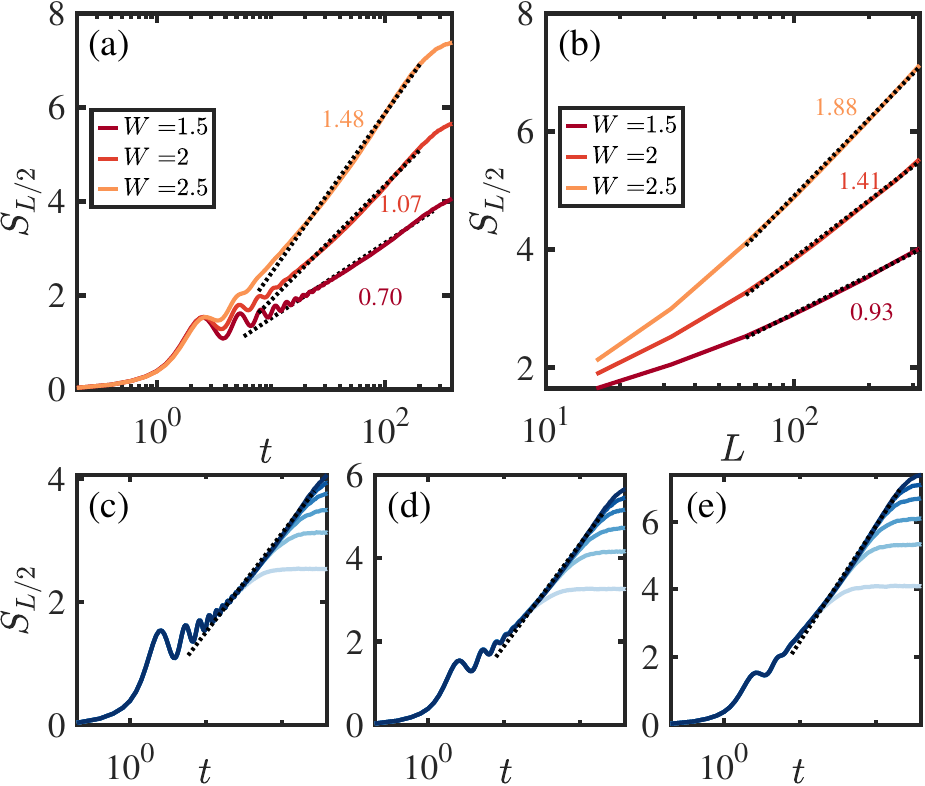}
\caption{Entanglement entropy under PBCs in the temporal and spatial directions, where the dashed lines are logarithmic fits.
(a) $S_{L/2}$ as a function of time $t$ for $L=384$.
(b) Steady-state entanglement entropy $S_{L/2}$ as a function of system size $L$.
(c--e) $S_{L/2}$ as a function of time $t$ for $W=1.5$, $2$, and $2.5$, respectively, with $L \in \{64,128,192,256,320,384\}$ from light to dark.}

\label{fig_spacetime_conformal}
\end{figure}

Boundary effects can significantly influence the properties of the steady states in non-Hermitian systems.
Under OBCs, the absence of conformal invariance in the steady state arises from the skin effects.
Due to the skin effects, particles move unidirectionally and eventually localize at the boundary, as illustrated in Fig.~2(b) in the main text.
Therefore, the localized particles exhibit little correlation with the bulk, as demonstrated in Fig.~2(c), where the correlation function decays faster than a power law away from the critical point.
This behavior contrasts with the predictions of conformal field theory, which suggests a power-law decay of the correlation function $C(x) \sim |x|^{-2 \Delta}$.
However, under PBCs, particles are allowed to circulate around the system without being localized.
The moving particles are able to build correlation and entanglement among different regions of the system.
Therefore, boundary effects can affect the steady states of non-Hermitian systems, especially for systems with skin effects.

To further investigate whether the system under PBCs exhibits spacetime conformal symmetry, we conduct numerical calculations of the entanglement entropy in both the spatial and temporal directions.
As shown in Fig.~\ref{fig_spacetime_conformal}, although the entanglement entropy exhibits logarithmic scaling in both directions, the prefactors are different.
According to Ref.~\cite{Lucas2020PRR}, the emergence of spacetime conformal symmetry in nonunitary dynamics requires identical prefactors for the logarithmic scaling in both spatial and temporal directions.
We thus conclude that our system does not exhibit full spacetime conformal symmetry.
Nevertheless, the steady states share certain qualitative features with conformal systems, such as power-law decaying correlation functions, nonzero mutual information, and logarithmic entanglement scaling.

\end{document}